\documentclass[10pt,twocolumn,letterpaper]{article}

\usepackage[pagenumbers]{cvpr} 
\usepackage{multirow}
\usepackage{graphicx}
\usepackage{multicol}
\usepackage{makecell}
\usepackage{booktabs} 
\usepackage{colortbl}
\usepackage{wasysym}

\usepackage[dvipsnames]{xcolor}

\definecolor{cvprblue}{rgb}{0.21,0.49,0.74}
\usepackage[pagebackref,breaklinks,colorlinks,citecolor=cvprblue]{hyperref}

\def\paperID{11110}
\def\confName{CVPR}
\def\confYear{2024}

\title{Modality-invariant and Specific Prompting for Multimodal Human Perception Understanding}

\author{Hao Sun\\
Zhejiang University\\
Hangzhou, China\\
{\tt\small sunhaoxx@zju.edu.cn}
\and
Ziwei Niu\\
Zhejiang University\\
Hangzhou, China\\
{\tt\small nzw@zju.edu.cn}
\and
Xinyao Yu\\
Zhejiang University\\
Hangzhou, China\\
{\tt\small xinyaoyu@zju.edu.cn}
\and
Jiaqing Liu\\
Ritsumeikan University\\
Shiga, Japan\\
{\tt\small liu-j@fc.ritsumei.ac.jp}
\and
Yen-Wei Chen\\
Ritsumeikan University\\
Shiga, Japan\\
{\tt\small chen@is.ritsumei.ac.jp}
\and
Lanfen Lin\\
Zhejiang University\\
Hangzhou, China\\
{\tt\small llf@zju.edu.cn}
}

\begin{document}
\maketitle
\begin{abstract}
Understanding human perceptions presents a formidable multimodal challenge for computers, encompassing aspects such as sentiment tendencies and sense of humor. While various methods have recently been introduced to extract modality-invariant and specific information from diverse modalities, with the goal of enhancing the efficacy of multimodal learning, few works emphasize this aspect in large language models. In this paper, we introduce a novel multimodal prompt strategy tailored for tuning large language models. Our method assesses the correlation among different modalities and isolates the modality-invariant and specific components, which are then utilized for prompt tuning. This approach enables large language models to efficiently and effectively assimilate information from various modalities. Furthermore, our strategy is designed with scalability in mind, allowing the integration of features from any modality into pretrained large language models. Experimental results on public datasets demonstrate that our proposed method significantly improves performance compared to previous methods.
\end{abstract}

\begin{figure}
    \centering
    \begin{minipage}{1.0\linewidth}
        \centering
        \includegraphics[width=1\textwidth]{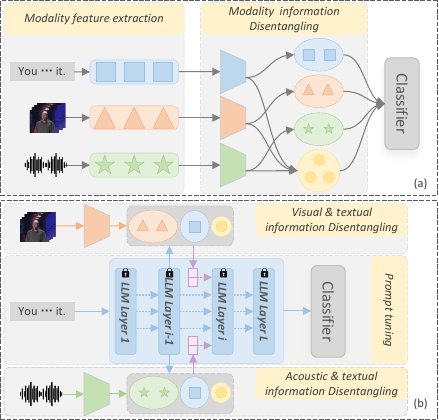}
    \end{minipage}
  \caption{The comparison of previous modality-invariant (\textcolor{yellow}{\CIRCLE}) and specific (\textcolor{orange}{$\Delta$} \textcolor{blue}{$\Box$} \textcolor{green}{$\bigstar$}) information learning(a) and our proposed multimodal prompting strategy(b), in which the modality-invariant and specific information are integrated.}
  \label{fig:first}
\end{figure}

\section{Introduction}
Understanding human perceptions, such as sentiment tendency and humor sense, encompasses a variety of emerging applications, such as online chatting~\cite{galik2012modelling,ray2023chatgpt} and dialogue systems~\cite{biswas2023role}. The primary challenge in this field lies in effectively leveraging human psychological information across various modalities, including text, acoustics, and facial cues.

In recent years, numerous approaches have emerged to decouple multimodal signals into modality-invariant and specific information for perception understanding~\cite{hazarika2020misa,yang2022disentangled} (as illustrated in Figure~\ref{fig:first}(a)). For instance, Hazarika et al.\cite{hazarika2020misa} decomposed involved modalities into several pieces and utilized central moment discrepancy\cite{zellinger2017central} to bring invariant parts closer while pushing specific parts further apart. These methods effectively capture human perception from multimodal features, remaining a key focus in multimodal research.

On another front, pretrained large language models (LLMs) have demonstrated excellent generalization abilities in various downstream tasks~\cite{yang2022re3,chan2022data}. However, due to their massive scale (e.g., 10 billion or 70 billion parameters), fine-tuning the entire model becomes challenging. Effectively prompting LLMs has thus become a prominent research topic in related fields~\cite{zhang2023llama,li2021prefix,liu2022few}. While many methods have endowed LLMs with multimodal processing capabilities~\cite{khattak2023maple,grounding2023}, fewer studies focus on enabling LLMs to leverage modality-invariant and specific information.

To address this gap, we propose a new multimodal prompting strategy with modality information disentangling for human perception estimation (as depicted in Figure~\ref{fig:first}(b)). Instead of directly extracting modality-invariant and specific information for final predictions, we emphasize these aspects in tunable prefix tokens. Our approach introduces the Parameter-Free Invariant and Specific prompt generation module (PaFIS) to generate tunable multimodal prompts with corresponding information from other modalities. Specifically, we calculate channel-wise correlations between text tokens and tokens from other modalities. We posit that less relevant parts contain more specific information, while more relevant parts bring more invariant information. By integrating these different parts into prompt tokens, LLMs can acquire the ability to process multimodal information. The entire process is parameter-free and does not introduce new learning parameters. Moreover, our strategy is designed with scalability, enabling the resolution of any modality features for LLM prompting. Experiments on four public datasets demonstrate that our approach significantly improves performance compared with other state-of-the-art methods. Through further analysis, our method also demonstrates the ability to handle modal absence cases. To the best of our knowledge, this is the first application that employs LLMs for modality information disentangling in human perception understanding.

In summary, our contributions can be summarized as follows:
\begin{itemize}
\item We propose a new multimodal prompting strategy, granting LLMs the ability to extract information from various modalities, such as facial and acoustic features.
\item We introduce PaFIS, a parameter-free modality-invariant and specific prompt generation module. Through PaFIS, LLMs can leverage modality-invariant and specific information from other modalities.
\item Our approach achieves state-of-the-art performance on four evaluated datasets, outperforming other methods by a large margin.
\end{itemize}

\section{Related Works}
Our work encompasses two primary research areas: multimodal learning and LLM tuning. Multimodal learning seeks to refine or discover relationships between multiple modalities, while LLM tuning aims to uncover the knowledge implicit in language models with efficient parameter approaches.

\subsection{Multimodal Representation Learning}
One research topic in multimodality focuses on learning effective representations from diverse modalities. Zadeh et al.\cite{zadeh2017tensor} introduced the Tensor Fusion Network (TFN), employing outer product to blend multimodal features into a compact representation. Gu et al.\cite{gu2018multimodal} proposed a hierarchical attention network to analyze multimodal signals. More recently, researchers have utilized the self-attention mechanism~\cite{vaswani2017attention} to fuse multimodal signals~\cite{tsai2019multimodal,delbrouck2020transformer,sun2023504}. Liu et al.\cite{liu2023osan} applied unsupervised domain adaptation to align involved modalities in a common space, and Sun et al.\cite{sun2022cubemlp} introduced a pure multilayer perceptron network for fusing multimodal features. Besides direct multimodal fusion, some researchers argue that modality-invariant and specific information are crucial for multimodal understanding. For instance, Hazarika et al.\cite{hazarika2020misa} used central moment discrepancy, while Yang et al.\cite{yang2022disentangled} employed Hilbert-Schmidt Independence Criterion~\cite{song2007supervised} to explicitly extract invariant and specific information for final predictions. These works highlight the importance of invariant and specific information in multimodal representation learning. Building upon this, our approach disentangles modality information with adaptability tailored for the fine-tuning of LLMs.

\subsection{LLMs for Language and Multimodal Learning}
In recent years, LLMs have gained significant attention in natural language processing and computer vision communities, demonstrating prowess in downstream tasks such as long-form generation and summarization~\cite{yang2022re3,chan2022data}. Most LLMs are based on the Transformer~\cite{vaswani2017attention} structure, boasting vast parameters and training data~\cite{zhang2022opt,touvron2023llama}. Directly tuning LLMs is infeasible due to their scale. Various approaches propose prompting large-scale models with few or no parameters, such as LoRA~\cite{hu2021lora}, IA3~\cite{liu2022few}, prefix tuning~\cite{li2021prefix}, and LLaMA-Adapter~\cite{zhang2023llama}. Researchers have also explored equipping LLMs with the ability to process information from other modalities. Merullo~\cite{merullo2022linearly} found similarities in latent representations between LLMs and large vision models (LVM), correlated through linear mappings. Jing et al.\cite{grounding2023} combined LLMs and LVMs with linear mappings for multimodal inputs and outputs. Khattak et al.\cite{khattak2023maple} used a projection function to map textual tokens to visual prompts. Our approach also adopts prompting for multimodal processing but places greater emphasis on extracting invariant and specific information from any other modalities.

\section{Methodology}
\begin{figure*}[ht]
\centering
\includegraphics[width=1\textwidth]{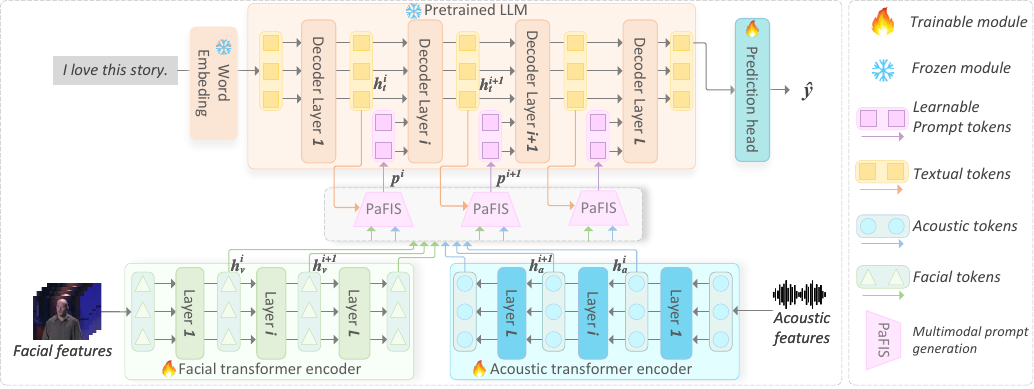}
\caption{The pipeline of our proposed method. The whole LLM is frozen while the facial and acoustic features are context modeled by respective transformer encoders. In each prompting layer $i$, textual($h_{t}^{i}$), facial($h_{v}^{i}$) and acoustic($h_{a}^{i}$) features are fed to Parameter-free invariant and specific prompt generation module(PaFIS). In PaFIS, the modality-invariant and specific information are integrated into learnable tokens $p^{i}$.}
\label{fig:pipeline}
\end{figure*}

Our approach concerns with fine-tuning a LLM with different kind of modalities while keeping its parameters frozen. As we focus on perception understanding, we mainly involve textual($t$), facial(visual, $v$), and acoustic($a$) features. We try to let the LLM leverage the modality-invariant and specific information during prompting, so as to further excavate the capabilities obtained from large scale pretraining.

\subsection{Model Architecture and Pipeline}
The framework overview is presented in Figure~\ref{fig:pipeline}. Text inputs are first tokenized and embedded before being fed into the pretrained LLM. The hidden states for the $i$th layer are represented as:
\begin{equation}
h_{t}^{i} = [h_{t, 1}^{i},h_{t, 2}^{i}, ... , h_{t, l_{t}}^{i}] \in \mathbb{R}^{l_{t}\times d_{t}},
\end{equation}
where $d_{t}$ is the embedding dimension, and $l_{t}$ is the number of tokens in the textual input. Simultaneously, facial and acoustic features undergo context modeling through respective Transformer encoders~\cite{vaswani2017attention}. The encoded states for the $i$th layer are expressed as:
\begin{equation}
h_{m}^{i} = [h_{m, 1}^{i},h_{m, 2}^{i}, ... , h_{m, l_{m}}^{i}] \in \mathbb{R}^{l_{m}\times d_{m}},
\end{equation}
where $m\in \{a, v\}$, $l_{m}$ is the sequential length of modality $m$, and $d_{m} (d_{m}<d_{t})$ is the corresponding feature dimension. To generate multimodal prompts for the $i$th layer of the LLM, we introduce a parameter-free invariant and specific prompt generation module (PaFIS):
\begin{equation}
p^{i} = \text{PaFIS}(h_{t}^{i}, h_{a}^{i}, h_{v}^{i}) \in \mathbb{R}^{l_{p}\times d_{t}},
\end{equation}
where $p^{i}$ is the prompt for the $i$th layer, and $l_{p}$ is the length of prompt tokens. Subsequently, we prefix the prompt tokens to the textual hidden states for the $i$th layer of the LLM:
\begin{equation}
h_{t}^{'i} = [p_{1}^{i}, ..., p_{l_{p}}^{i}; h_{t, 1}^{i}, ... , h_{t, l_{t}}^{i}] \in \mathbb{R}^{(l_{p}+l_{t})\times d_{t}}.
\end{equation}
During processing, we maintain frozen LLM parameters and exclusively train the facial and acoustic Transformers for context modeling.

\subsection{Parameter-free Invariant and Specific Prompt Generation(PaFIS)}
The PaFIS module aims to generate learnable prompts where modality-invariant and specific information are emphasized. In the $i$th layer of the LLM, the PaFIS module takes modality hidden states ($h_{t}^{i}$, $h_{a}^{i}$, $h_{v}^{i}$) as inputs and outputs multimodal prompts $p^{i}$. For each PaFIS module, we first introduce the learnable prompts $\tilde{p}^{i}\in \mathbb{R}^{l_{p}\times d_{t}}$, which are not only used to tune the LLM but also serve as containers for modality-invariant and specific information.

\textbf{Aligned Cases.} 
When acoustic or visual features are strictly aligned with textual tokens ($l_{m}=l_{t}=l$), we calculate the channel-wise correlation coefficients\footnote{The Pearson Correlation Coefficient is employed in our work.} between $h_{m}^{i}$ and distinct channels of $h_{t}^{i}$:
\begin{equation}
\begin{aligned}
K = \text{Corr}(h_{m}^{i}, h_{t}^{i}[j:j+d_{m}])\in & \mathbb{R}^{l\times (d_t-d_m)},\\
\text{for } j \in & [0, \ d_{t}-d_{m}],
\end{aligned}
\end{equation}
where $h_{t}^{i}[j:j+d_{m}]\in \mathbb{R}^{l\times d_m}$ is the sub-channels of $h_{t}$ used to calculate coefficients with $h_{m}^{i}$. Next, we identify the distinct sub-channels of $h_{t}^{i}$ with the highest correlation:
\begin{equation}
\begin{aligned}
k_{m}^{max} &= \text{argmax}(K, \text{axis}=1) \in \mathbb{N}^{l},  \\
c_{m}^{max} &= [k_{m}^{max}[j]:k_{m}^{max}[j]+d_{m}] \in \mathbb{N}^{l\times d_m}, j \in [0, l], \\
h_{t}^{i,inv} &= h_{t}^{i}[c_{m}^{max}] \in \mathbb{R}^{l\times d_m}.
\end{aligned}
\end{equation}
Similarly, we identify those with the lowest correlation:
\begin{equation}
\begin{aligned}
k_{m}^{min} &=  \text{argmin}(K, \text{axis}=1) \in \mathbb{N}^{l}, \\
c_{m}^{min} &= [k_{m}^{min}[j]:k_{m}^{min}[j]+d_{m}] \in \mathbb{N}^{l\times d_m}, j \in [0, l], \\
h_{t}^{i,spe} &= h_{t}^{i}[c_{m}^{min}] \in \mathbb{R}^{l\times d_m}.
\end{aligned}
\end{equation}
We posit that $h_{t}^{i,inv}$ channels (of high correlation with $h_{m}^{i}$) bring more modality-invariant information, while $h_{t}^{i,spe}$ channels (of low correlation with $h_{m}^{i}$) contain more modality-specific information. To enable the LLM to leverage this information from other modalities, we integrate them into the learnable prompts:
\begin{equation}
\begin{aligned}
\label{equ:address}
p^{i}[c_{m}^{max}] &= \tilde{p}^{i}[c_{m}^{max}]+(h_{m}^{i}+h_{t}^{i,inv}), \\
p^{i}[c_{m}^{min}] &= \tilde{p}^{i}[c_{m}^{min}]+(h_{m}^{i}-h_{t}^{i,spe}).
\end{aligned}
\end{equation}
In Equation~\ref{equ:address}, modality-invariant information(between modality $t$ and $m$) is emphasized by $(h_{m}^{i}+h_{t}^{i,inv})$ while specific information(from modality $m$) is emphasized by $(h_{m}^{i}-h_{t}^{i,spe})$.

\textbf{Unaligned Cases.} In cases where acoustic or facial features are not aligned with textual tokens, we average-pool the hidden states on the temporal level before calculating the correlation:
\begin{equation}
\begin{aligned}
K = \text{Corr}(\bar{h}_{m}^{i}, \bar{h}_{t}^{i}[j:j+d_{m}])\in \mathbb{R}^{(d_t-d_m)},& \\
\text{for } j \in [0, \ d_{t}-d_{m}].&
\end{aligned}
\end{equation}
where $\bar{h}_{m}^{i}$ and $\bar{h}_{t}^{i}$ are temporally pooled features. Correspondingly, the sub-channels of $\bar{h}_{t}^{i}$ with the highest correlation are calculated by:
\begin{equation}
\begin{aligned}
k_{m}^{max} &= argmax(K) \in \mathbb{N}, \\
c_{m}^{max} &= [k_{m}^{max}:k_{m}^{max}+d_{m}] \in \mathbb{N}^{d_m},\\
\bar{h}_{t}^{i,inv} &= \bar{h}_{t}^{i}[c_{m}^{max}] \in \mathbb{R}^{d_m},
\end{aligned}
\end{equation}
The same is true for the lowest correlated channels $\bar{h}_{t}^{i,spe}$. Likewise, modality-invariant and specific information will be emphasized in learnable prompts:
\begin{equation}
\begin{aligned}
p^{i}[j][c_{m}^{max}] = \tilde{p}^{i}[j][c_{m}^{max}]+(\bar{h}_{m}^{i}+\bar{h}_{t}^{i,inv}), &\\
p^{i}[j][c_{m}^{min}] = \tilde{p}^{i}[j][c_{m}^{min}]+(\bar{h}_{m}^{i}-\bar{h}_{t}^{i,spe}), &\\
\text{for } \ j \in [0, \ l_{t}]&.
\end{aligned}
\end{equation}

\begin{table*}[ht]
\centering
\caption{The results of our method on MOSI and MOSEI dataset. $\downarrow$: the lower the better; $\uparrow$: the higher the better. *: previous state-of-the-art performance.}
\label{tab:resultmosi}
\begin{tabular}{c|cccc|cccc}
\toprule
& \multicolumn{4}{c|}{MOSI} & \multicolumn{4}{c}{MOSEI} \\
& MAE($\downarrow$) & Corr($\uparrow$) & Acc-2/F1($\uparrow$) & Acc-7($\uparrow$) & MAE($\downarrow$) & Corr($\uparrow$) & Acc-2/F1($\uparrow$) & Acc-7($\uparrow$) \\
\midrule
TFN\cite{zadeh2017tensor}     & 0.970 & 0.633 & 73.9/73.4 & 32.1 & 0.593 & 0.700 & 82.5/82.1 & 50.2 \\
ICCN\cite{sun2020learning}    & 0.862 & 0.714 & 83.0/83.0 & 39.0 & 0.565 & 0.713 & 84.2/84.2 & 51.6 \\
MulT\cite{tsai2019multimodal} & 0.871 & 0.698 & 83.0/82.8 & 40.0 & 0.580 & 0.703 & 82.5/82.3 & 51.8 \\
MISA\cite{tsai2019multimodal} & 0.817 & 0.748 & 82.1/82.0 & 41.4 & 0.557 & 0.748 & 84.9/84.8 & 51.7 \\
BBFN\cite{han2021bi}          & 0.776 & 0.775 & 84.3/84.3 & 45.0 & 0.529 & 0.767 & 86.2/86.1*& 54.8*\\
CMLP\cite{sun2022cubemlp}     & 0.770 & 0.767 & 85.6/85.5*& 45.5 & 0.529 & 0.760 & 85.1/84.5 & 54.9 \\
MMIM\cite{han2021improving}   & 0.700*& 0.800 & 84.1/84.0 & 46.6*& 0.526*& 0.772*& 82.2/82.6 & 54.2 \\
OSAN\cite{liu2023osan}        & 0.713 & 0.801*& 83.1/83.0 & 46.3 & 0.532 & 0.768 & 83.4/83.3 & 53.8 \\
\midrule
\rowcolor[HTML]{EFEFEF}
Ours & \textbf{0.619} & \textbf{0.860} & \textbf{86.5/86.5} & \textbf{49.3} & \textbf{0.501} &  \textbf{0.789} & \textbf{87.3/87.2} & \textbf{55.4}\\
\rowcolor[HTML]{EFEFEF}
$\Delta_{SOTA}$ & $\downarrow$11.6\% & $\uparrow$7.4\% & $\uparrow$0.9\%/1.0\% & $\uparrow$2.7\% & $\downarrow$4.8\% & $\uparrow$2.2\% & $\uparrow$1.1\%/1.1\% & $\uparrow$0.5\%\\
\bottomrule
\end{tabular}
\end{table*}

\begin{table*}[ht]
\centering
    \caption{The results of our method on POM dataset. $\downarrow$: the lower the better; $\uparrow$: the higher the better. \textit{Con, Pas, Voi, Dom, Cre, Viv, Exp, Ent, Per, Res, Tru, Rel, Out, Tho, Ner, Hum, Laz, Sen} mean for \textit{confident, passionate, voice pleasant, dominant, credible, vivid, expertise, entertaining, persuasion, reserved, trusting, relaxed, outgoing, thorough, nervous, humorous, laziness, sentiment}, respectively. $\Diamond$ means that the metrics are not provided in presented papers, we rerun the methods based on public codes. *: previous SOTA method.} 
    \label{tab:resultpom}
    
    \begin{subtable}{1.0\linewidth}
      \centering
        \caption{Task name with MAE metric.}
        \renewcommand\arraystretch{1.2}
        \setlength{\tabcolsep}{0.8mm}{}
        \resizebox{1.0\textwidth}{!}{
        \begin{tabular}{@{}ccccccccccccccccccc@{}}
        \toprule
        \multicolumn{1}{c|}{\multirow{2}{*}{Methods}} & \multicolumn{18}{c}{Task Name [MAE($\downarrow$) metric]} \\
        \multicolumn{1}{c|}{}& Con & Pas & Voi & Dom & Cre & Viv & Exp & Ent & Per & Res & Tru & Rel & Out & Tho & Ner & Hum & Laz & Sen\\
        \midrule
        \multicolumn{1}{c|}{MFN\cite{zadeh2018memory}$_\Diamond$}  & 0.952 & 0.993 & 0.882 & 0.835 & 0.903 & 0.908 & 0.886 & 0.913 & 0.981 & 0.821 & 0.521 & 0.566 & 0.679 & 0.665 & 0.654 & 0.727 & 0.911 & 0.742 \\
        \multicolumn{1}{c|}{MulT\cite{tsai2019multimodal}$_\Diamond^*$} & 0.844 & 0.879 & 0.791 & 0.766 & 0.804 & 0.877 & 0.830 & 0.900 & 0.852 & 0.745 & 0.500 & 0.638 & 0.622 & 0.641 & 0.634 & 0.747 & 0.857 & 0.761\\
        \midrule
        \rowcolor[HTML]{EFEFEF}
        \multicolumn{1}{c|}{Ours} & \textbf{0.810} & \textbf{0.821} & \textbf{0.731} & \textbf{0.722} & \textbf{0.792} & \textbf{0.783} & \textbf{0.777} & \textbf{0.821} & \textbf{0.801} & \textbf{0.699} & \textbf{0.443} & \textbf{0.500} & \textbf{0.527} & \textbf{0.592} & \textbf{0.510} & \textbf{0.603} & \textbf{0.731} & \textbf{0.736} \\
        \rowcolor[HTML]{EFEFEF}
        \multicolumn{1}{c|}{$\Delta_{SOTA}(\downarrow)$ } & 4.0\% & 6.6\% & 7.6\% & 5.7\% & 1.5\% & 10.7\% & 6.4\% & 8.8\% & 6.0\% & 6.2\% & 11.4\% & 21.6\% & 15.3\% & 7.6\% & 19.6\% & 17.1\% & 14.7\% & 0.8\% \\
        \bottomrule \\
        \end{tabular}}
    \end{subtable}
    
    \begin{subtable}{0.48\linewidth}
      \centering
        \caption{9 POM tasks with Acc-7 metric.}
        \renewcommand\arraystretch{1.2}
        \setlength{\tabcolsep}{0.9mm}{}
        \resizebox{0.96\textwidth}{!}{
        \begin{tabular}{@{}c|ccccccccc@{}}
        \toprule
        \multirow{2}{*}{Methods} & \multicolumn{9}{c}{Task Name [Acc-7($\uparrow$) metric]} \\
        & Con & Pas & Voi & Dom & Cre & Viv & Exp & Ent & Sen \\
        \midrule
        SVM             & 26.6 & 20.7 & -    & 35.0 & 25.1 & -    & -    & 31.5 & -    \\
        TFN\cite{zadeh2017tensor}$_\Diamond$  & 24.1 & 31.0 & 38.7 & 34.5 & 24.6 & 39.9 & 32.0 & 29.1 & 41.1 \\
        MFN\cite{zadeh2018memory}$_\Diamond$  & 34.5 & 35.5 & 37.4 & 41.9 & 34.5 & 36.9 & 36.0 & 37.9 & 52.0 \\
        MARN\cite{zadeh2018multi}  & 29.1 & 33.0 & -    & 38.4 & 31.5 & -    & -    & 33.5 & -    \\
        MulT\cite{tsai2019multimodal}$_\Diamond^*$ & 39.0 & 40.2 & 43.6 & 39.2 & 40.5 & 38.8 & 43.1 & 40.9 & 57.8 \\
        \midrule
        \rowcolor[HTML]{EFEFEF}
        Ours            & \textbf{43.1} & \textbf{44.2} & \textbf{45.1} & \textbf{43.9} & \textbf{44.5} & \textbf{47.0} & \textbf{44.5} & \textbf{43.8} & \textbf{69.2} \\
        \rowcolor[HTML]{EFEFEF}
        $\Delta_{SOTA}(\uparrow)$ & 4.1\% & 4.0\% & 1.5\% & 4.7\% & 4.0\% & 8.2\% & 1.4\% & 2.9\% & 11.4\% \\
        \bottomrule 
        \end{tabular}}
    \end{subtable}
    \begin{subtable}{0.48\linewidth}
      \centering
        \caption{9 POM tasks with Acc-5 metric.}
        \renewcommand\arraystretch{1.2}
        \setlength{\tabcolsep}{0.9mm}{}
        \resizebox{0.96\textwidth}{!}{
        \begin{tabular}{@{}c|ccccccccc@{}}
        \toprule
        \multirow{2}{*}{Methods} & \multicolumn{8}{c}{Task Name [Acc-5($\uparrow$) metric]} &  \\
        & Per & Res & Tru & Rel & Out & Tho & Ner & Hum & Laz \\
        \midrule
        SVM             & 28.1 & 34.0 & 50.2 & 49.8 & -     & -     & 41.4 & 36.0 & - \\
        TFN\cite{zadeh2017tensor} $_\Diamond$  & 27.6 & 30.5 & 38.9 & 35.5 & 44.3  & 50.5  & 42.4 & 33.0 & 38.4 \\
        MFN\cite{zadeh2018memory} & 34.0 & 38.4 & 57.1 & 53.2 & 46.8  & 47.3  & 47.8 & 47.3 & 37.2 \\
        MARN\cite{zadeh2018multi} & 31.0 & 36.9 & 55.7 & 52.2 & -     & -     & 47.3 & 44.8 & - \\
        MulT\cite{tsai2019multimodal}$_\Diamond^*$ & 37.8 & 46.6 & 57.1 & 53.0 & 54.2 & 56.4  & 50.9 & 48.8 & 41.3 \\
        \midrule
        \rowcolor[HTML]{EFEFEF}
        Ours            & \textbf{42.9} & \textbf{48.4} & \textbf{62.1} & \textbf{60.1} & \textbf{55.4}  & \textbf{58.6}  & \textbf{52.5} & \textbf{51.7} & \textbf{46.3} \\
        \rowcolor[HTML]{EFEFEF}
        $\Delta_{SOTA}(\uparrow)$  & 5.1\% & 1.8\% & 5.0\% & 6.9\% & 1.2\% & 2.2\% & 1.6\% & 2.9\% & 5.0\%\\
        \bottomrule 
        \end{tabular}}
    \end{subtable}
    
\end{table*}

\begin{table}[t]
\caption{The results of our method on URFunny dataset. $\uparrow$: the higher the better. \textit{Pre} means the precision and \textit{Rec} means the recall. $\Diamond$: some of the metrics are not provided by presented papers, we rerun the methods based on public codes. *: SOTA performance.}
\label{tab:result_urfunny}
\centering
\begin{tabular}{c|ccc}
\toprule
\multirow{2}{*}{Models} & \multicolumn{3}{c}{URFunny} \\
& Acc-2/F1($\uparrow$) & Pre($\uparrow$) & Rec($\uparrow$) \\
\hline
TFN\cite{zadeh2017tensor}$_\Diamond$       & 68.5/68.6 & 67.9 & 61.1 \\
MulT\cite{tsai2019multimodal}$_\Diamond$   & 70.5/70.4 & 66.4 & 70.0 \\
MISA\cite{tsai2019multimodal}$_\Diamond$   & 70.6/70.0 & 68.9 & 73.5* \\
BBFN\cite{han2021bi}$_\Diamond$            & 71.6/72.0* & 69.2* & 72.8 \\
DRL\cite{yang2022disentangled} & 71.8/ - & - & - \\
\midrule
\rowcolor[HTML]{EFEFEF}
Ours & \textbf{74.2}/\textbf{74.2} & \textbf{71.8} & \textbf{77.7} \\
\rowcolor[HTML]{EFEFEF}
$\Delta_{SOTA}(\uparrow)$ & 2.6\%/2.2\% & 2.6\% & 4.2\% \\
\bottomrule
\end{tabular}
\end{table}

\subsection{Training Target}
Our framework addresses two types of tasks: regression and binary classification. However, LLMs are typically designed and pretrained with language generation tasks. Therefore, we append a fully-connected layer after the last token hidden state $h_{t,l_{t}}^{L}$ in the last layer for predictions. For regression tasks, we utilize the rooted mean square error as the loss function:
\begin{equation}
\mathcal{L}=\sqrt{\frac{1}{N}\sum_{i=1}^{N}(y_n-\hat{y}_n)^2},
\end{equation}
where $N$ is the number of samples in a training batch, $y_n$ is the ground truth, and $\hat{y}_n$ is the prediction. For classification tasks, the binary cross-entropy loss is used:
\begin{equation}
\mathcal{L}=\frac{1}{N}\sum_{i=1}^{N}[y_n\log \hat{y}_n+(1-y_n)\log (1-\hat{y}_n)].
\end{equation}

\section{Experiments}
\subsection{Datasets Evaluated}
To assess the effectiveness of our proposed method, we conducted experiments on four publicly available datasets: MOSI~\cite{zadeh2016multimodal}, MOSEI~\cite{zadeh2018multimodal}, URFunny~\cite{hasan2019ur}, and POM~\cite{park2014computational}. Each of these datasets provides textual, facial, and acoustic modalities.

\textbf{MOSI \& MOSEI:} These datasets consist of utterances gathered from a public social platform. MOSI comprises 1283 training utterances, 229 validation utterances, and 686 testing utterances. In MOSEI, there are 16315 training samples, 1817 validation samples, and 4654 testing samples. Labels range from $[-3, +3]$, where -3 and +3 denote the most negative and positive sentiments, respectively. Although we approach the task as a regression, we can also provide 2-class and 7-class metrics by rounding the labels.

\textbf{URFunny:} This dataset is designed for humor detection in video clips, with clips sourced from the Internet and annotated as 0 or 1 based on the presence of humor punchlines. It consists of 10598 training samples, 2626 validation samples, and 3290 testing samples.

\textbf{POM:} The POM dataset includes 600 training samples, 100 validation samples, and 203 testing samples. Each sample offers 18 human perceptions, covering sentiment, confidence, passion, voice pleasantness, dominance, credibility, vividness, expertise, entertainment, reserve, trust, laziness, relaxation, extroversion, thoroughness, nervousness, humor, and persuasion. All labels fall within the range of $[1, 7]$, where a higher value indicates a stronger tendency.

\subsection{Experimental settings}
In our experiments, we employed the popular LLaMA2-7B~\cite{touvron2023llama} as the foundational model, which stands as one of the state-of-the-art open-source LLMs in the natural language processing community. Inherited from LLaMA2, $d_{t}$ is set to 4096, and $d_{m}$ varies depending on the type of features but is consistently set to be less than 128. For performance reasons, we fixed the prompt token length $l_{p}$ to 8 and only prefixed the prompts in the last three layers of the LLM. Moreover, to adapt to the text generation task of LLM pre-training, we formulate the original text input as \textit{`Below is a text that describes a movie. Predict the \{task\} according to the text. \#\#\# Text: \{text\}  \#\#\#\{task\} tendency:'}
where \{text\} is the input sentence, \{task\} varies according to different tasks, e.g. sentiment, humor, confidence.

The learning rate was set to 0.0001, and the batch size was set to 8 on each GPU. We trained the model on datasets for up to 40 epochs with early stopping. All methods were implemented using the PyTorch framework and mixed-precision with bfloat16~\cite{abadi2016tensorflow}. Our experiments were conducted on two Nvidia RTX3090Ti GPUs, taking less than an hour when training on the MOSI dataset. \footnote{The codes for our method will be made public after publication.}

\begin{table*}[t]
\caption{The ablation studies on MOSI and URFunny dataset. $h_a$ and $h_v$ means whether we use the acoustic or facial modality features. PaFIS means whether the modality-invariant and specific information are addressed during tuning. Test$a$ or Test$v$ mean whether to use acoustic or visual features during inference and testing.}
\label{tab:ablation}
\centering
\begin{tabular}{ccc|cc|cccc|cc}
\toprule
\multicolumn{5}{c|}{Ablations} & \multicolumn{4}{c|}{MOSI} & \multicolumn{2}{c}{URFunny} \\
\hline
PaFIS & $h_a$ & $h_v$ & Test$a$ & Test$v$ & MAE($\downarrow$) & Corr($\uparrow$) & Acc-2/F1($\uparrow$) & Acc-7($\uparrow$) & Acc-2($\uparrow$) & F1($\uparrow$) \\
\hline
           &             &            &            &            & 0.689 & 0.840 & 82.2/82.2 & 46.9 & 72.1 & 72.1 \\
\checkmark & \checkmark  &            & \checkmark &            & 0.639 & 0.844 & 83.7/83.7 & 48.2 & 73.0 & 72.9 \\
\checkmark &             & \checkmark &            & \checkmark & 0.644 & 0.841 & 82.2/83.9 & 47.1 & 71.1 & 71.1 \\
           &  \checkmark & \checkmark & \checkmark & \checkmark & 0.630 & 0.843 & 84.4/84.3 & 47.9 & 72.9 & 71.4 \\
\hline
\rowcolor[HTML]{EFEFEF}
\checkmark & \checkmark & \checkmark & & &0.655 & 0.820 & 82.8/82.6 & 45.4 & 71.8 & 71.8 \\
\rowcolor[HTML]{EFEFEF}
\checkmark & \checkmark & \checkmark & \checkmark & & 0.638 & 0.839 & 83.4/83.5 & 46.9 & 73.4 & 73.3 \\
\rowcolor[HTML]{EFEFEF}
\checkmark & \checkmark & \checkmark & & \checkmark & 0.641 & 0.830 & 83.0/83.0 & 46.1 & 72.6 & 72.5 \\
\rowcolor[HTML]{EFEFEF}
\checkmark & \checkmark & \checkmark & \checkmark & \checkmark & \textbf{0.619} & \textbf{0.860} & \textbf{86.5/86.5} & \textbf{49.3} & \textbf{74.2} & \textbf{74.2} \\
\bottomrule
\end{tabular}
\end{table*}

\section{Results and Analysis}
Our results, along with comparisons to other approaches, are presented in Table~\ref{tab:resultmosi}, Table~\ref{tab:resultpom}, and Table~\ref{tab:result_urfunny}. We benchmark our method against current state-of-the-art approaches, including Tensor Fusion Network~\cite{zadeh2017tensor}, MISA~\cite{hazarika2020misa}, and OSAN~\cite{liu2023osan}. As evident from the tables, our method significantly outperforms the current state-of-the-art approaches. Specifically, we achieve a Mean Absolute Error (MAE) of 0.619 on the MOSI dataset and 0.501 on the MOSEI dataset for sentiment tendency prediction. In humor detection, we attain an accuracy of 74.2 on the URFunny dataset. Furthermore, our method achieves state-of-the-art performance across all 18 tasks on the POM dataset.

\begin{figure}
    \centering
    \begin{minipage}{1.0\linewidth}
        \centering
        \includegraphics[width=1\textwidth]{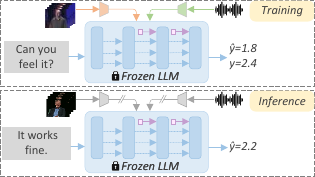}
    \end{minipage}
  \caption{The illustration of modality absence inference(the case in figure is for sentiment analysis). The model is tuned in with multimodal inputs(up) but without multimodal signals for inference(down).}
  \label{fig:noavinput}
\end{figure}

\subsection{Ablation and Modality Robustness}
To further assess the effectiveness of our approach, we conducted ablation studies on the MOSI and URFunny datasets. Our focus was on studying the impact of the PaFIS module and the involved modalities. In scenarios where the PaFIS module is not employed, the modality features $h_{a}^{i}$ and $h_{v}^{i}$ are directly added to the textual hidden states $h_{t}^{i}$ at arbitrary positions, implying that modality-invariant and specific information is not emphasized. The results are presented in Table~\ref{tab:ablation}. The observed performance drop without the PaFIS module emphasizes the crucial role of modality-invariant and specific information in multimodal learning. Through the metrics, we observe that the enhancement in our performance is not solely attributed to LLM; however, the contributions of PaFIS and other modalities should not be underestimated.

Furthermore, we explored the significance of different modalities in human perception estimation. As evident from the results, both acoustic and facial information ($h_{a}$, $h_{v}$) play crucial roles in perceptual understanding. However, acoustic features tend to contribute more to the predictions than facial features. Additionally, we experimented with fine-tuning the LLM with full modalities but only utilizing texts during inference (as illustrated in Figure~\ref{fig:noavinput}). Surprisingly, the performance (shown in Figure~\ref{tab:ablation} with Test$_{a}$/Test$_v$ ablation) remained superior to previous work, as illustrated in Table~\ref{tab:resultmosi} and Table~\ref{tab:result_urfunny}. This indicates the robustness of our proposed multimodal prompting strategy in scenarios where modalities are absent, which is common in real-time applications.

\subsection{Cross Dataset Evaluation}
As some datasets provide the same kind of acoustic and facial features, we conducted cross-dataset evaluation experiments. Two models were trained on the MOSEI dataset: model1 was trained with COVAREP~\cite{degottex2014covarep} ($a$) and Facet~\cite{ekman1997face} ($v$) features, while model2 was trained with COVAREP~\cite{degottex2014covarep} ($a$) and OpenSMILE~\cite{eyben2010opensmile} ($v$) features. After training, the models were directly tested on the MOSI and POM datasets, which provide the same type of features as the two models above, respectively. As indicated in Table~\ref{tab:crossdataset}, the results remained competitive with previous approaches. This suggests that our proposed approach facilitates generalization capabilities in perception understanding.

\begin{table}[t]
\centering
    \caption{The results of cross dataset evaluation on MOSI and POM dataset. On POM dataset, we only evaluate on the sentiment analysis task.} 
    \label{tab:crossdataset}
    
    \begin{subtable}{1.0\linewidth}
      \centering
        \caption{The results of our method on MOSI dataset when training on MOSEI.}
        \resizebox{1.0\textwidth}{!}{
        \begin{tabular}{ccccc}
        \toprule
        \multicolumn{1}{c|}{Methods} & MAE($\downarrow$) & Corr($\uparrow$) & Acc-2/F1($\uparrow$) & Acc-7($\uparrow$) \\
        \midrule
        \multicolumn{1}{c|}{MMIM\cite{han2021improving}} & \textbf{0.700} & 0.800 & \textbf{84.1/84.0} & \textbf{46.6} \\
        \multicolumn{1}{c|}{OSAN\cite{liu2023osan}} & 0.713 & 0.801 & 83.1/83.0 & 46.3 \\
        \midrule
        \rowcolor[HTML]{EFEFEF}
        \multicolumn{1}{c|}{Ours} & 0.719 & \textbf{0.802} & 83.2/83.3 & 46.4 \\
        \bottomrule \\
        \end{tabular}
        }
    \end{subtable}

    \begin{subtable}{1.0\linewidth}
      \centering
        \caption{The results of our method on POM dataset(sentiment regression task) when training on MOSEI.}
        \begin{tabular}{c|cc}
        \toprule
        Method & MAE($\downarrow$) & Acc-7($\uparrow$) \\
        \midrule
        MFN\cite{zadeh2018memory} & \textbf{0.742} & 52.0 \\
        MulT\cite{tsai2019multimodal} & 0.761 & 57.8 \\
        \midrule
        \rowcolor[HTML]{EFEFEF}
        Ours & 0.759 & \textbf{60.1} \\
        \bottomrule
        \end{tabular}
    \end{subtable}
    
\end{table}

\begin{figure}
    \centering
    \begin{minipage}{1.0\linewidth}
        \centering
        \includegraphics[width=1\textwidth]{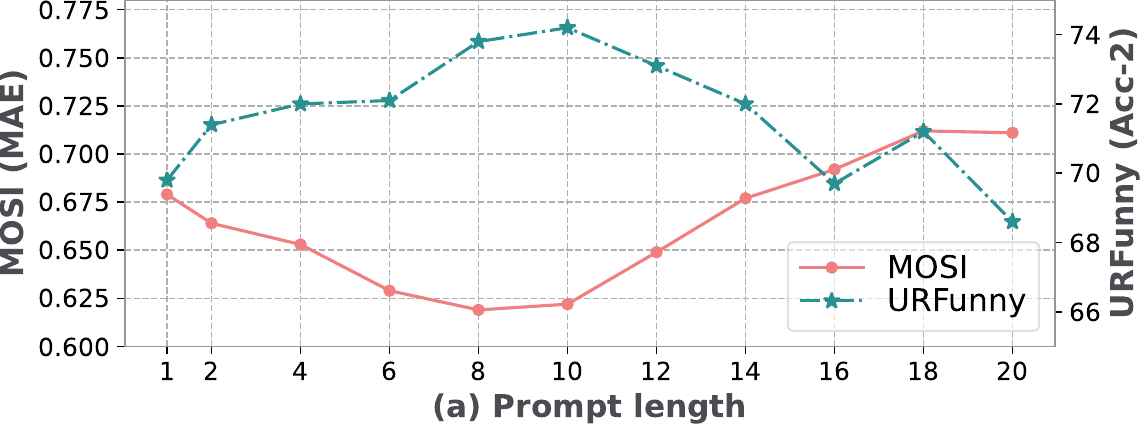}
    \end{minipage}
    \begin{minipage}{1.0\linewidth}
        \centering
        \includegraphics[width=1\textwidth]{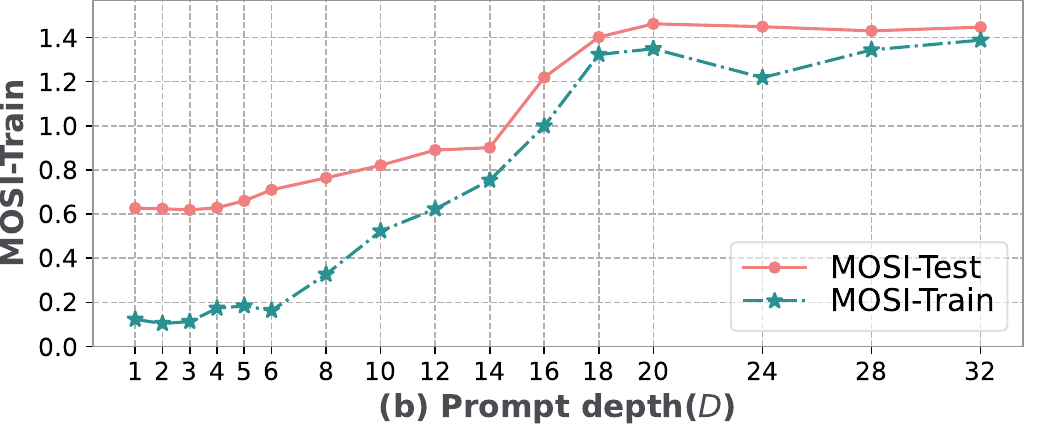}
    \end{minipage}
  \caption{The impact of prompt scales to the final predictions. (a) The impact of prompt length. (b) The impact of prompt depth($D$).}
  \label{fig:prompscale}
\end{figure}

\subsection{Impact of Prompt Scales}
As the LLM is frozen when performing multimodal prompting, the prompts play a crucial role in the final performance. Therefore, we studied the influence of prompt scales, including prompt length and prompt depth.

\textbf{Prompt Length:} Figure~\ref{fig:prompscale}(a) illustrates the effect of prompt length on our proposed method. The experiments were conducted on the MOSI and URFunny datasets. Longer prompt lengths indicate stronger fitting ability but come with a greater computational burden and a risk of overfitting. We observed that the best results were achieved with a prompt length ($l_{p}$) set between 8 and 10.

\textbf{Prompt Depth:} In Figure~\ref{fig:prompscale}(b), we showcase the effect of different prompt depths on predictions. In LLM, we consistently prompt the last $D$ layers, as deeper features contain richer semantic information and are more conducive to multimodal learning. The experiments were conducted on the MOSI dataset. Results demonstrate that performance is optimal when the prompt depth is set to 3. Any increase or decrease in the depth of prompts results in reduced performance. An increase in prompt depth leads to overfitting, while shallower prompts can lead to underfitting. We present not only the performance on the test set but also the results on the training set. Interestingly, the model fails to fit on the training set when prompts are deeper than 15 ($D>15$). We attribute this to the scale of data in downstream tasks. A surplus of tunable parameters relative to the data scale eventually hampers model fitting.

\begin{table}[t]
\centering
    \caption{The impact of different encoders for acoustic and facial encoders. The experiments are conducted on MOSI dataset.} 
    \label{tab:avencoders}
    \resizebox{1.0\linewidth}{!}{
    \begin{tabular}{c|cccc}
    \toprule
    Encoders & MAE($\downarrow$) & Corr($\uparrow$) & Acc-2/F1($\uparrow$) & Acc-7($\uparrow$) \\
    \midrule
    Convolution & 0.710 & 0.839 & 81.8/81.7 & 46.7 \\
    LSTM & 0.625 & 0.851 & 85.2/85.1 & 47.8 \\
    GRU & 0.622 & 0.849 & 85.0/85.1 & 48.0 \\
    \midrule
    \rowcolor[HTML]{EFEFEF}
    Transformer & \textbf{0.619} & \textbf{0.860} & \textbf{86.5/86.5} & \textbf{49.3} \\
    \bottomrule
    \end{tabular}
    }
\end{table}

\subsection{Selection of Acoustic \& Facial Encoders}
As evident from the ablation studies presented in Table~\ref{tab:ablation}, acoustic and facial features play a significant role in perception understanding. Consequently, we investigated the impact of different encoders for modality features other than texts. Four types of encoders were employed: LSTM~\cite{hochreiter1997long}, GRU~\cite{cho2014learning}, convolution, and Transformer encoders~\cite{vaswani2017attention}. The findings in Table~\ref{tab:avencoders} indicate that Transformer yields the best results, while LSTM and GRU exhibit slightly inferior performance. Interestingly, the convolution encoders produce the poorest results, even worse than the text-only model in Table~\ref{tab:ablation}. This suggests that convolution not only fails to extract effective information but also, at times, has a negative impact on LLM.

\section{Conclusion}
This paper introduced a novel multimodal prompting strategy for Large Language Models (LLMs), leveraging the Parameter-Free Invariant and Specific prompt generation module (PaFIS). The PaFIS module generates learnable prompts that integrate modality-invariant and specific information, enabling LLMs to effectively utilize information from various modalities. Experimental results on four public datasets, including MOSI, MOSEI, URFunny, and POM, showcased the effectiveness of our proposed method. Despite these positive outcomes, our approach is susceptible to overfitting due to the potent representation capability of LLMs. Future research will focus on refining strategies to seamlessly integrate multimodal signals into LLMs while addressing overfitting concerns.

\section*{Acknowledgments}
The acknowledgments and sponsors are hidden for double-blind review.


{
\small
\bibliographystyle{ieeenat_fullname}
\bibliography{references}
}


\end{document}